# Suppressed dependence of polarization on epitaxial strain in highly polar ferroelectrics

Ho Nyung Lee,[1,*] Serge M. Nakhmanson,[2,**] Matthew F. Chisholm,[1] Hans M. Christen,[1] Karin M. Rabe,[2] and David Vanderbilt[2]

[1]Materials Science and Technology Division, Oak Ridge National Laboratory, Oak Ridge, Tennessee 37831

[2]Department of Physics and Astronomy, Rutgers University, Piscataway, New Jersey 08854-8019

A combined experimental and computational investigation of coupling between polarization and epitaxial strain in highly polar ferroelectric PbZr$_{0.2}$Ti$_{0.8}$O$_3$ (PZT) thin films is reported. A comparison of the properties of relaxed (tetragonality $c/a \approx 1.05$) and highly-strained ($c/a \approx 1.09$) epitaxial films shows that polarization, while being amongst the highest reported for PZT or PbTiO$_3$ in either film or bulk forms ($P_r \approx 82$ μC/cm$^2$), is almost independent of the epitaxial strain. We attribute this behavior to a suppressed sensitivity of the A-site cations to epitaxial strain in these Pb-based perovskites, where the ferroelectric displacements are already large, contrary to the case of less polar perovskites, such as BaTiO$_3$. In the latter case, the A-site cation (Ba) and equatorial oxygen displacements can lead to substantial polarization increases.

Epitaxial strain, induced in thin films due to lattice mismatch between the material and the substrate, results in enhanced properties and device performance for many materials. Examples include a higher operation speed and lower power consumption in strain-engineered semiconductor-based devices [1] and a large enhancement of ferroelectric and dielectric responses in certain complex oxide perovskites [2–6]. Therefore, epitaxial strain is recognized as a useful tool to influence materials properties and it is tempting to assume that such strong sensitivity to epitaxial strain is common to all perovskite compounds.

However, recent computational studies suggest that this is not universally true [7], while an experimental confirmation of such nonuniversality is a formidable task due to limitations explained below. Moreover, the fundamental mechanisms responsible for the coupling between epitaxial strain and ferroelectric polarization in thin films are not yet fully understood, although it is quite clear that they should be manifested most profoundly in coherently grown thin films. In such films misfit dislocation formation is energetically and/or kinetically prohibited and therefore reduction of the polarization by pinning of dipoles by defects is unlikely.

This is, however, true only for films with thicknesses below a certain critical value ($t_c$), above which the strain relaxation via defect formation occurs. This thickness depends on the lattice mismatch between the film and the substrate [8,9]: for example, in BaTiO$_3$ on SrTiO$_3$ (in-plane mismatch of 2.4%) a $t_c \approx$ 2-4 nm has been experimentally determined [10,11]. But at such small film thicknesses polarization measurements are hindered by high tunnelling and/or leakage currents. A strong influence of depolarization fields, originating from the incomplete screening of polarization [12], and a reduction of atomic displacements in the vicinity of interfaces [13] may also mask the intrinsic effects. Therefore, a careful characterization of strain effects is most reliably achieved in a system with a fairly small lattice mismatch with the substrate and, consequently, large $t_c$ (e.g., several tens of nanometres). Similar considerations have already led to successful studies of strain effects in SrTiO$_3$ [Ref. 4] and BaTiO$_3$ [Ref. 3].

In this work, we grow PbZr$_{0.2}$Ti$_{0.8}$O$_3$ films (PZT, $a = b = $ 0. 3953 nm, $c = $ 0.4148 nm, $c/a \approx$ 1.05 for an unstrained film) [14] by pulsed laser deposition (PLD), which results in a 1.2 % lattice mismatch with the SrTiO$_3$ substrate (cubic with $a = $ 0.3905 nm). For the lattice-matched bottom electrodes on the SrTiO$_3$ substrates, we use atomically-flat, epitaxially-strained SrRuO$_3$ films [5,15]. To elucidate the effect of epitaxial strain on the magnitude of polarization, the structural and ferroelectric properties of PZT films are investigated both experimentally and by first-principles calculations. The latter includes comparisons with other similar perovskite materials, such as BaTiO$_3$ and PbTiO$_3$. We show, in both cases, that the polarization enhancement by biaxial compressive strain induced in thin films of *highly polar ferroelectrics* is unexpectedly suppressed.

The results shown in Fig. 1 demonstrate the structural quality of the current PZT thin films grown on SrRuO$_3$ (4 nm) on (001) SrTiO$_3$ substrate. X-ray diffraction (XRD) $\theta$–$2\theta$ scans confirm the *c*-axis orientation of the PZT films without any contribution from *a*-axis oriented domains and impurity phases. SrRuO$_3$ bottom electrodes (same as in Ref. 5) are utilized as templates to grow the PZT films with smooth surfaces, as evidenced by the finite thickness fringes in the XRD scans [Fig. 1(a)]. While only the out-of-plane lattice constant, $c$, is determined by the diffraction peaks in XRD $\theta$–$2\theta$ scans, the average in-plane and out-of-plane lattice constants are simultaneously determined by XRD reciprocal space maps (not shown). The maps confirm that the films grow closely lattice-matched on SrRuO$_3$-electroded SrTiO$_3$ up to about 40 nm in thickness ($a \approx 0.391$ nm, $c/a \approx 1.09$). Above 40 nm PZT thickness, the films relax to a bulk-like state ($a \approx 0.395$ nm, $c/a \approx 1.05$) as seen in Fig. 1(b). The sudden decrease of tetragonality in films thicker than ~40 nm suggests that the kinetic barrier for strain relaxation is exceeded at this film thickness. This relatively large critical thickness is attributed to the low growth temperature (625 °C), consistent with the reported behavior in Ge$_x$Si$_{1-x}$ [Ref. 9]. However, there is still a small but non-negligible drop in the $c/a$ ratio from 1.095 to 1.087 as the film thickness increases from 15 nm to 40 nm [Fig. 1(b)]. This is related to an increase in the number of dislocations, which do not significantly contribute to the strain relaxation due to their limited number, e.g. the average dislocation spacing seen in a cross section of the 15 nm thick film [Fig. 1(c)] is ~170 nm. A cross section from a relaxed film (116 nm thick), on the other hand, shows a larger number of dislocation lines with an average spacing of ~30 nm [Fig. 1(d)]. This is indeed in good agreement with the fact that an orthogonal array of dislocations (Burgers vector: **b** = 0.3905 nm) with a spacing of 32 nm would completely relieve the misfit strain. Figure 2a shows $P(E)$ hysteresis loops for two PZT films, with thicknesses of 30 nm (strained) and 100 nm (relaxed). Both films exhibit fully saturated, well-defined square hysteresis loops without any indication of the polarization relaxation. The obtained values for the remnant polarization, $P_r \approx 82$ μC/cm$^2$ measured at 100 Hz, are amongst the highest reported for PZT and PbTiO$_3$ in film or bulk forms. Pulse measurements of polarization further confirm the absence of any appreciable leakage contribution [switching polarization: $P_{sw}$ = 162.3 μC/cm$^2$ and non-switching polarization: $P_{nsw}$= 0.4 μC/cm$^2$ even from the 35 nm-thick film as shown in Fig. 2(b)]. The well-defined switching current curves in the inset of Fig. 2(b) also indicate that the high values obtained for the polarization are not a consequence of instrumental artefacts or high leakage current. Surprisingly, as shown in Figs. 2(c) and (d), our polarization data reveal almost no difference between the values in strained and relaxed films, despite the fact that epitaxial strain increases the tetragonal distortion ($c/a$ -1) by almost a factor of two (from ~5% in the relaxed films to ~9% in the strained films). Such a lack of polarization enhancement is quite striking in a material praised for its "strong tetragonal strain-polarization coupling" and high piezoelectric coefficients [16,17].

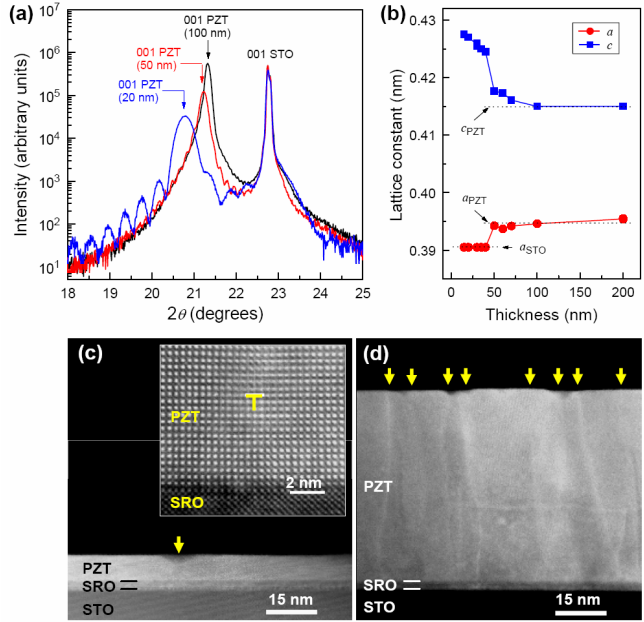

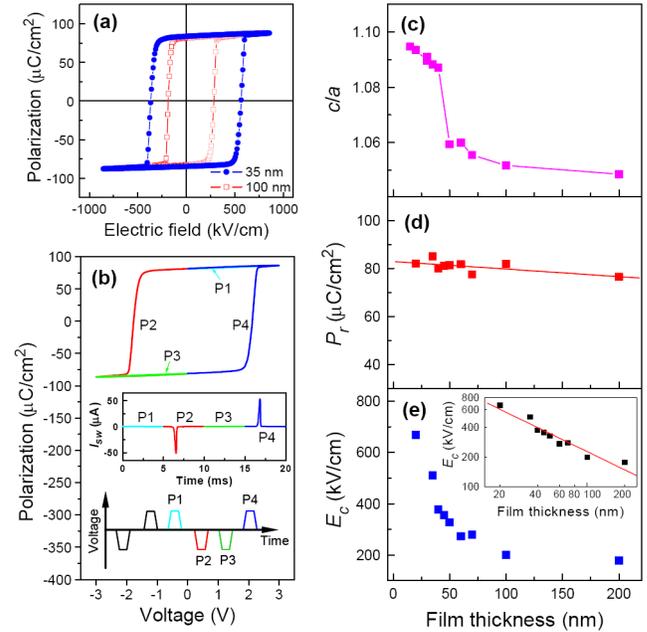

Figure 1 (Color online) (a) XRD $\theta$–$2\theta$ scans around the 001 reflection of 20, 50, and 100 nm thick films, showing the increase of out-of-plane lattice constant by reducing the film thickness. Note that the reflection from SrRuO$_3$ is not appreciable due to a rather small thickness (4 nm). (b) In-plane and out-of-plane lattice constants as a function of film thickness measured from XRD scans. Cross-sectional Z-contrast images of (c) 15 and (d) 116 nm thick PZT films. The inset of (c) is a high-resolution image of the 15 nm thick film showing a dislocation core located 6 nm above the SrRuO$_3$ layer. The solid arrows in (c) and (d) point to the dislocation lines, indicating the reduced dislocation density in the well-strained film. STO and SRO stand for SrTiO$_3$ and SrRuO$_3$, respectively.

Figure 2 (Color online) (a) $P(E)$ hysteresis loops at 100 Hz from 35 and 100 nm thick PZT films. (b) Simultaneously-measured pulse polarization and switching current (inset) by applying trapezoid pulses (P1 to P4), showing neither non-switching current nor non-switching polarization. Pulse measurements in (b) were performed at 3 V with 2.5 ms of write/read-pulse time and one second write-pulse delay. Tetragonality (c), remanent polarization (d), and coercive field (e) as a function of film thickness. The slope (–0.614±0.067) in the inset of (e) is in good agreement with the classical scaling behaviour ($E_c \propto d^{-2/3}$).

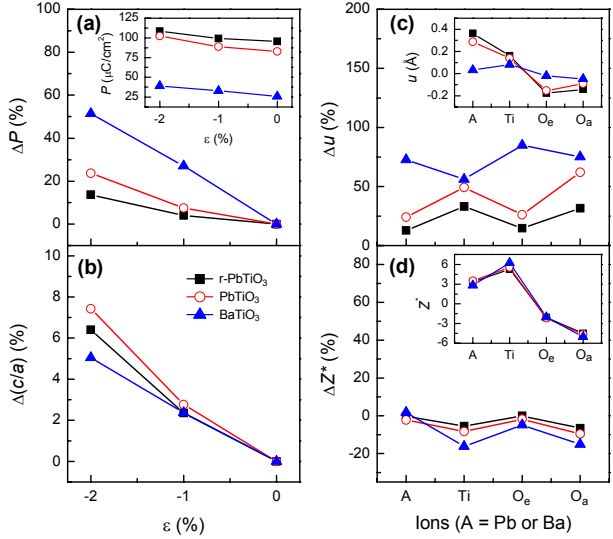

Figure 3 (Color online) Polarization enhancement (a), tetragonality enhancement (b), and polarization [inset in (a)] as a function of epitaxial strain $\varepsilon$ in r-PbTiO$_3$, PbTiO$_3$, and BaTiO$_3$. The enhancement of $u$ and $Z^*$ by epitaxial strain $\varepsilon = -2\%$ is shown in (c) and (d), respectively. Insets in (c) and (d) (same $x$ axes as main figures) are the values of $u$ and $Z^*$, respectively, at zero strain. O$_e$ and O$_a$ refer to equatorial and apical oxygens, respectively.

In addition to the changes of film tetragonality and polarization [Figs. 2(c) and (d)], it is also instructive to analyze the behavior of the coercive field $E_c$ (i.e., the field required to reverse the polarization) as a function of film thickness. Results extracted from hysteresis loops similar to the ones shown in Fig. 2(a) are summarized in Fig. 2(e). While the values of the coercive fields are significantly larger than those reported for single crystals (10 ~ 100 kV/cm), the dependence follows the Kay-Dunn scaling law ($E_c(d) \propto d^{-2/3}$) [18] over the entire film thickness range (20 nm – 200 nm), consistent with previous reports [19], and is undisturbed by the abrupt change of tetragonality at around 50 nm. We can therefore conclude that the Kay-Dunn law is insensitive to the presence of bulk defects or the epitaxial-strain state of the films, but may still reflect the changes in switching kinetics related to domain nucleation/growth or depolarization fields with varying film thickness [20].

To elucidate the microscopic phenomena behind the observed modest coupling between polarization and epitaxial strain in thin PZT films, we have complemented the experiments with a first principles investigation of atomistic PZT models [21]. Since the experimental 20/80 Zr-to-Ti concentration ratio is difficult to reproduce in a compact unit cell, it was changed to a more accommodating 25/75 ratio. By averaging over the supercell family, we obtained $P = 87.3$ μC/cm$^2$ and $c/a = 1.061$ for the relaxed PZT, and $P = 92.6$ μC/cm$^2$ and $c/a = 1.086$ for the commensurately strained PZT, which constitutes a 6% polarization increase for an averaged in-plane strain $\varepsilon = -1\%$. These results are in excellent agreement with the experimental observations of the absence of substantial polarization enhancement by epitaxial strain.

In order to further understand this phenomenon, we have calculated the enhancement of $P$ and $c/a$ as a function of epitaxial strain for tetragonal BaTiO$_3$ and PbTiO$_3$ as shown in Figs. 3(a) and (b). Additionally, a "rescaled PbTiO$_3$" (r-PbTiO$_3$) material – with in-plane lattice constant tuned by hand to match the tetragonality of the Ti-rich PZT – has been examined, providing a convenient reference system that mimics certain properties of the disordered PZT solid solution. For $\varepsilon$ changing from 0 to –2% (bracketing the experimental value of –1.2%), all three structures show a similar increase in tetragonality and polarization: $\Delta c/a \approx$ 5-7% and $P(\varepsilon) - P(0) \approx$ 15-20 μC/cm$^2$. However, since the zero-strain polarization in the lead-based structures is much higher than in BaTiO$_3$, the polarization enhancement $\Delta P(\%) = 100 \cdot [P(\varepsilon) - P(0)] / P(0)$ in the former appears to be quite small, compared to the latter material.

To obtain detailed information about the polarization change, we employed a linearized approximation that splits the out-of-plane polarization into a set of ionic contributions [2,6]: $P \approx V^{-1} \sum_i Z_i^* u_i$. Here $V$ is the cell volume, while $Z_i^*$ and $u_i$ are, respectively, the effective charge tensor component and the ferroelectric displacement of ion $i$ in the out-of-plane direction. As shown in Figs. 3(c) and (d), the effective charges of the same ionic types for all materials are similar and decrease slightly under applied strain, while the ionic displacements in unstrained BaTiO$_3$ are 2-5 times smaller than in PbTiO$_3$ and r-PbTiO$_3$, leading to a smaller total polarization. When the strain is applied ($\varepsilon = -2\%$), the ionic displacements in BaTiO$_3$ (especially for Ba and O$_e$) increase more dramatically than in PbTiO$_3$ and r-PbTiO$_3$, resulting in a much greater enhancement of polarization.

Overall, it can be then concluded that since unstrained PbTiO$_3$ and PZT are already very polar, and ferroelectric ionic displacements in them (especially on Pb) are very large, they are rather insensitive to applied epitaxial strain. In other words, while the Pb-O hybridization results in strong lattice distortions that are associated with large polarization [16], these displacements – and thus the polarization – cannot be substantially enlarged by applying additional strain. The less polar BaTiO$_3$, on the other hand, is much more sensitive to strain, with the largest changes occurring for the A-site (Ba) and O$_e$ ions that have the smallest out-of-plane displacements in the unstrained configuration, typical of B-site driven ferroelectrics [25]. We can thus tie our experimental observations directly to the magnitude of ferroelectric ion off-centering, without excluding, however, a possible additional contribution of stereochemically active lone pairs in these Pb-based perovskites to the suppressed dependence of polarization on epitaxial strain, similar to observations reported for BiFeO$_3$ [7].

In summary, epitaxially-strained and -relaxed PZT thin films have been grown by PLD. We found that, below the critical thickness (*ca.* ~40 nm), the tetragonality in thin films is indeed significantly increased (to a value of $c/a = 1.095$) by biaxial compressive strain, as compared to that in relaxed films ($c/a = 1.05$). However, the polarization values $P_r \approx 80$ μC/cm$^2$ recorded from epitaxially-strained films are almost identical to those from relaxed films, despite the fact that the tetragonal distortion is increased by almost a factor of two by epitaxial strain. Consistently, our first-principles study confirms not only the experimental observation of the weak dependence of polarization on epitaxial strain in such highly polar materials, but also that the polarization dependence on epitaxial strain varies considerably for different ferroelectrics. The computational (atomic-scale) insights into the nature of the interplay between polarization and epitaxial strain in perovskite-type ferroelectrics show further that the ionic sensitivity to epitaxial strain is less pronounced in highly polar materials.

The authors thank D. H. Lowndes and D. J. Singh for useful discussion. Research sponsored by the Division of Materials Sciences and Engineering, Office of Basic Energy Sciences, U.S. Department of Energy, under contract DE-AC05-00OR22725 with Oak Ridge National Laboratory, managed and operated by UT-Battelle, LLC and by the Center for Piezoelectrics by Design (CPD) under ONR Grant N00014-01-1-0365. The calculations were done at MHPCC DoD supercomputing center.

*e-mail: hnlee@ornl.gov
**Present address: *Materials Sciences Division, Argonne National Laboratory, Argonne, IL 60439; supported by the UChicago Argonne, LLC, under Contract #DE-AC02-06CH11357*